\documentstyle[aps,twocolumn]{revtex}

\input epsf
\begin{document}
\draft
\title{Towards a global classification of excitable reaction--diffusion systems}
\author{Silvina Ponce Dawson~\cite{email}~$^{1}$, Mar\'\i a Ver\'onica D'Angelo~$^1$ and
John E. Pearson~\cite{emailj}$^2$}

\address{$^1$ Departamento de F\'\i sica \\
Facultad de Ciencias Exactas y
Naturales, U.B.A. \\
Ciudad Universitaria, Pabell\'on I \\
(1428) Buenos Aires, Argentina}

\address{$^2$ Applied Theoretical and Computational Physics\\ Los Alamos
National Laboratory\\ XCM MS F645\\ Los Alamos, NM 87545, USA
}

\maketitle

\begin{abstract}
Patterns in reaction--diffusion systems near primary bifurcations 
can be studied locally and classified by means of amplitude equations. 
This is not possible for excitable reaction--diffusion systems. 
In this Letter we propose a global classification of two variable
excitable reaction--diffusion systems. In particular, we claim that
the topology of the underlying two--dimensional homogeneous dynamics
can be used to organize the system's behavior.

\end{abstract}


Many dissimilar experimental and model reaction--diffusion systems
display similar behavior.  This raises the question of whether a
classification can be found so that seemingly unrelated systems with
similar dynamics can be understood as belonging to the same
equivalence class. We argue in this Letter that a global
classification is possible for reaction--diffusion systems whenever
the underlying homogeneous dynamics can be mapped onto a {\it planar
flow} ({\it i.e.}, a flow in $R^2$).  This classification is done in
terms of {\it model families}~\cite{cross} that are largely determined
by the homogeneous dynamics.  In support of our conjecture we analyze
experiments done in an open reactor using the FIS
(Ferrocyanide--Iodate--Sulfite) reaction~\cite{swinney,exp-spots} and
two reaction--diffusion models~\cite{john,muratov} that display
similar patterns to those of the experiment, even though they are not
accurate models of the FIS kinetics.  We explain these common
behaviors by noting that they all have similar homogeneous dynamics
and discuss the main features of their model family.

The classification of system behaviors lies at the heart of dynamical
systems theory and here the normal form theorem is one of the
main achievements~\cite{gh}. It states that close to a bifurcation
point the dynamics of any sufficiently smooth system can be reduced to
a simplified set of equations that is locally topologically equivalent
to the full vector field~\cite{gabo}.  The normal form approach is
local in both phase and parameter space.  Thus, classifications based
on it provide generic descriptions only when the system is near
threshold.  For this reason, the patterns that occur in the {\it
excitable systems} that we discuss in this Letter cannot be classified with
normal form techniques.

Excitability is a common dynamical behavior that occurs, for example,
in the kinetics of neuronal membrane potentials~\cite{fitzhugh60}, in
semiconductor lasers with feedback~\cite{manuel} and in a variety of
chemical reactions~\cite{swinney}.  The picture of excitability we use
is defined in terms of the spatially uniform dynamics. It assumes the
existence of a stable fixed point (a spatially uniform stationary
solution) such that perturbations above a threshold result in a large
excursion in phase space before the systems decays back to the fixed
point.  A spatially uniform excitable dynamics can result in wave
propagation or in multistability when spatial variations are allowed.
In particular this is true when the coupling is diffusive, {\it i.e.},
for {\it reaction--diffusion} systems (see {\it e.g.}~\cite{john}).
Since these behaviors arise from a finite perturbation of the
spatially uniform steady state, classifications based on local normal
forms cannot be used.

We now consider a reaction--diffusion system whose homogeneous
dynamics can be mapped onto a planar flow.  Because of the limited set
of asymptotic behaviors allowed by flows in $R^2$~\cite{2d} they are
particularly easy to classify. Given the fixed points of the flows
there is a finite set of possible asymptotic behaviors and transitions
among them~\cite{aclar2}.  Thus, we can classify {\it families} of
planar flows that smoothly deform into one another by varying
parameters~\cite{future}.  It is important to note that this is a {\it
global} description, as opposed to the local one provided by near
threshold normal forms. This global description is valid for the
spatially uniform dynamics.  When diffusion terms are added the planar
system becomes high--dimensional. We conjecture that there
exists a natural global description for the high--dimensional
reaction--diffusion system that is the direct analogue of the known
global description of the two--dimensional planar system.  This means
there should exist a {\it model family} of two coupled
reaction--diffusion equations which displays the same patterns and
transitions among them as any other member of the class.  The central
statement is that, for reaction--diffusion systems which in the
homogeneous limit can be described by planar flows, these model
families are mainly determined by the homogenous dynamics.
 
Given a set of two reaction--diffusion equations with bounded
diffusion terms, the local dynamics at each time and point in space
is, properly understood, a perturbation of the homogeneous planar
flow. 
Thus, the model family of the class which a particular system belongs to should not only
display homogeneous behaviors that are topologically equivalent to those of the system
we are trying to classify, but it should also contain its closest perturbations. 
 Since
flows in $R^2$ with a finite number of fixed points have a finite set
of asymptotic behaviors and can undergo a countable set of
bifurcations, we then expect to find a countable set of model
families. This gives the desired classification of excitable
systems describable by pairs of reaction--diffusion equations.  In
this Letter we present a particular application which supports our
viewpoint.

\begin{figure}
\epsfxsize=6 true cm
\hskip 0.7 true cm \epsfbox{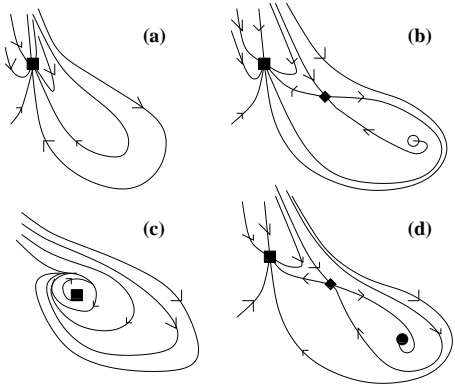}
\vskip 0.5 true cm 
\caption{ Planar flow with one 
excitable fixed point (a) and its closest perturbations (b)--(d).
}
\label{fig:exc}
\end{figure}

We first develop a simple geometric model of excitability in $R^2$
suitable for our purposes. We consider a set of equations $\dot
u=f(u,v)$, $\dot v=g(u,v)$, with only one attractor: a stable fixed
point, $\overline{x}\equiv (\overline{u},\overline{v})$.  Excitability
implies that the flow lines turn around before coming back to
$\overline{x}$.  The simplest situation with this property is depicted
in Fig.~\ref{fig:exc}~(a), that corresponds to the case in which
$\overline{x}$ (the square) is the only limit set of the flow.  This
flow can generically undergo a ``saddle--repellor'', a saddle--node or
a Hopf bifurcation, leading to the flows shown in
Figs.~\ref{fig:exc}~(b)--(d), respectively. The fixed point
$\overline{x}$ is stable in all cases but in (d), where there is an
attracting limit cycle.  In Fig.~\ref{fig:exc}~(b) and (c) it coexists
with a saddle (the diamond) and a repellor (the white circle) or a
node (the black circle).  While the Hopf bifurcation requires that
$\overline{x}$ be a spiral, which is unrelated to it being excitable,
the saddle--node and saddle--repellor bifurcations may be linked to
the excitability of $\overline{x}$. In fact, excitability is related
to the existence of a separatrix, which becomes the stable manifold of
the saddle after it is born.  Also, the turn around of the flow lines
implies that the orientation of the {\it nullclines} (the curves that
satisfy $f(u,v)=0=g(u,v)$) is such that, by not too large a
perturbation, they can eventually become tangent at a point.
Generically, when this happens, a saddle--node or saddle--repellor
bifurcation occurs. These two types of bifurcations ``meet'' at a
codimension two point, the Takens--Bogdanov point.  In fact, the
simplest family that contains the flows in
Figs.~\ref{fig:exc}~(a)--(c) outside a neighborhood of $\overline{x}$
is given by the Takens--Bogdanov normal form~\cite{tak}:
%\begin{eqnarray}
$\dot u = v$,
$\dot v = \mu_1 + \mu_2 v + u^2 + uv$.
%\label{eq:tak_bog}
%\end{eqnarray}
It is easy to modify these equations so that they also describe the
behavior at and near $\overline{x}$ (see {\it e.g.}~\cite{manuel}).
%In any case, even if we cannot call such a family
%a normal form in the usual sense, it does contain all possible behaviors
%and bifurcations of flows in $R^2$ with one stable
%fixed point and at most two more fixed points and one 
%periodic orbit. 
Flows of this ``extended'' family  with only one fixed point 
have the inflection of the flow lines necessary for excitability.
We may thus call it a general model of excitability in $R^2$, for systems
with one fixed point. Since the family contains all the relevant perturbations
of the flow in Fig.~\ref{fig:exc}~(a), it is also
suitable in the extended case~\cite{aclar_pert}.
\begin{figure}
\epsfxsize=6 true cm
\hskip 0.7 true cm \epsfbox{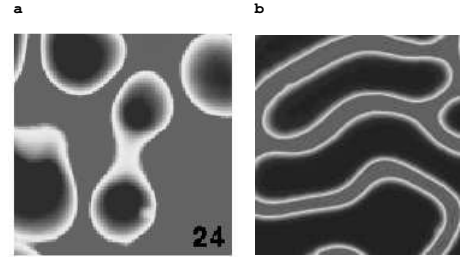}
\vskip 0.5 true cm 
\caption{Transition from replicating spots to labyrinthine patterns
in the FIS reaction when
the homogeneous system approaches the saddle--node saddle--repellor bifurcation.
}
\label{fig:exp}
\end{figure}

We now consider  a reaction--diffusion system with an underlying
planar homogeneous dynamics
that has only one stable but excitable fixed point.
As a model family of its class we choose a two--variable
reaction--diffusion system that, in the homogeneous limit,
contains the flows in Figs.~\ref{fig:exc}~(a)--(c).  Now, any affine
transformation $(u,v)\rightarrow {\cal M} (u-u_0,v-v_0)$, with $\cal
M$ a constant $2\times 2$ matrix, will give another family with an
equivalent homogeneous dynamics and, in many cases, the right
``winding'' of the flow lines.  If we introduce such a transformation in
the set $\partial _t u= D_u \nabla ^2 u + f(u,v)$, $\partial _t v= D_v
\nabla ^2 v + g(u,v)$, we get a new set with cross-diffusion terms.
Thus, once we have any family, $\dot u= f(u,v)$, $\dot v= g(u,v)$,
with the ``right'' homogeneous dynamics, we expect, in most cases,
that each (stationary) pattern of the system of interest be equivalent
to a pattern of a flow in the family
\begin{eqnarray}
\partial _t u&=&D_{uu}\nabla^2 u + D_{uv}\nabla^2 v+f(u,v),\nonumber\\ 
\partial _t v&=&D_{vu}\nabla^2 u + D_{vv}\nabla^2 v+g(u,v).
\label{eq:gen}
\end{eqnarray} 
Since it is possible to get rid of cross--diffusion by a simple transformation,
it should always be possible to choose a {\it particular} 
model family without it.

Now we discuss three concrete examples of systems with the excitable
homogeneous dynamics discussed above. 
Consider the FIS reaction, which produces the patterns of
Fig.~\ref{fig:exp}~\cite{exp-spots}.  In the well--mixed ({\it i.e.},
spatially homogeneous) case,
it is known to exhibit excitability, bistability and
oscillations~\cite{fis_exp}.  These homogeneous behaviors can be
described by planar dynamical systems. The dynamical models
of the system (a set of 10 ODE's)~\cite{fis_theo} show that for
the parameter values used in all the experiments, there is a
separation of timescales 
\begin{figure}
\epsfxsize=1.6truein
\hskip -0.7 true cm \epsfbox{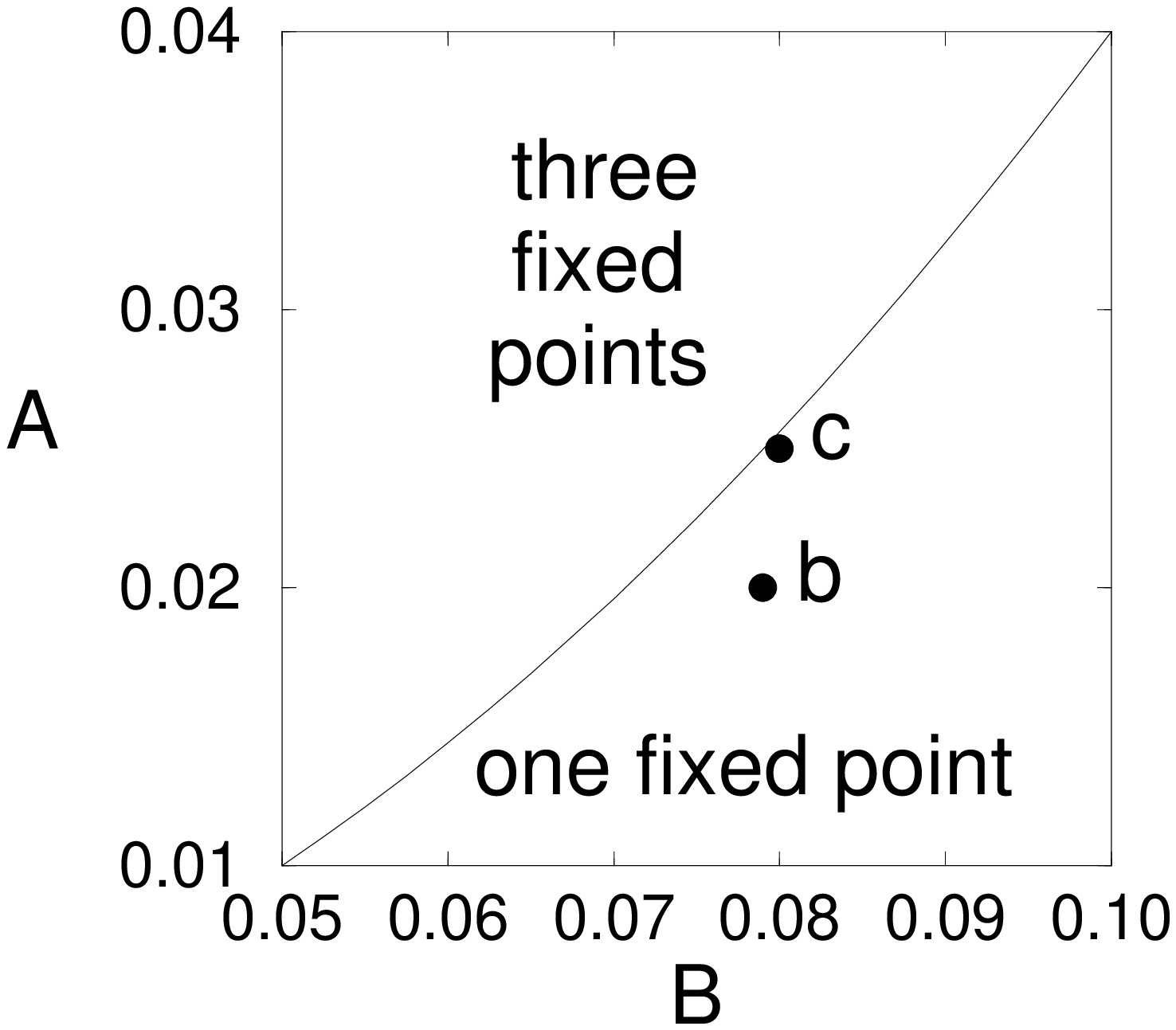}
\epsfxsize=1.3truein
\vskip -3 true cm\hskip  2.5 true cm\epsfbox{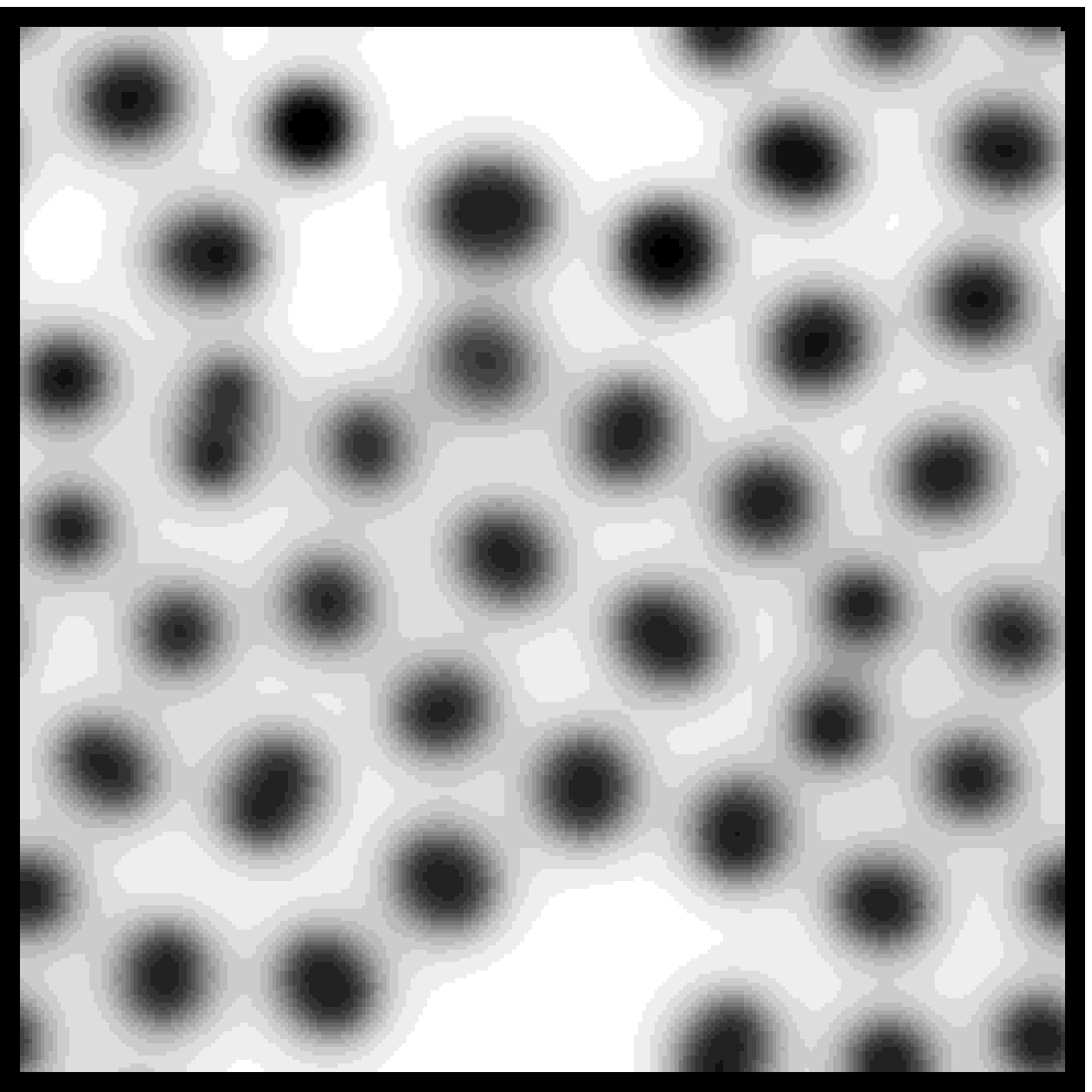}
\epsfxsize=1.3truein
\vskip -2.4 true cm\hskip  4.99 true cm\epsfbox{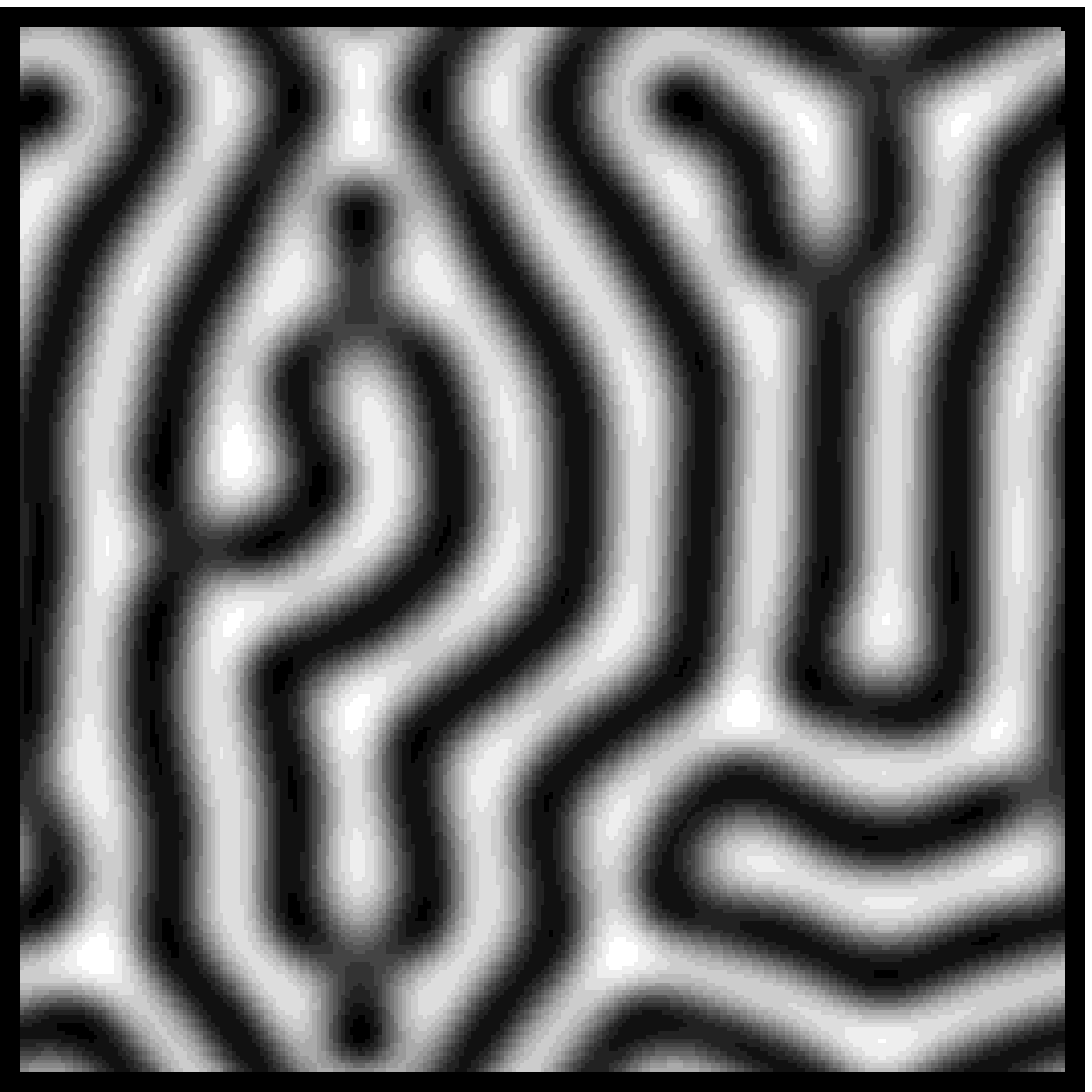}
%\end{figure}
\vskip -2.8 cm
\hskip 1.6 true cm
%\fontsize{14}{16pt}\selectfont
a
\hskip 2.2 cm
b
\hskip 2.2 cm
c
\vskip 3 true cm\caption{ A partial bifurcation set for the homogeneous Gray--Scott
model (a) and two examples of patterns that occur for the parameters
indicated in (a) when $D_u=1$ and $D_v=0.5$.  The line corresponds to the curve of
saddle-repellor bifurcations.}
\label{fig:gs}
\end{figure}
that allows the reduction of the original
10--dimensional system to a planar one. This planar system can have
at most three fixed points, one of them a saddle. Experimentally,
only stable solutions can be observed. The existence of the unstable
fixed points and other limit sets are deduced from the model.

In this Letter we discuss the replicating spots
(Fig.~\ref{fig:exp}~(a)) and lamellar structures
(Fig.~\ref{fig:exp}~(b)) that are found when there is only one
homogeneous stationary solution, the low pH fixed point, and it is stable~\cite{exp-spots}. 
Experimentally, the spots are initiated by
finite perturbations of the low pH state. This reflects its
excitability.  Thus, a model family like the one sketched before should
reproduce at least some of the observed patterns.  The transition from
spots to lamellae shown in Fig.~\ref{fig:exp} is observed as the
concentration of ferrocyanide, $[{\rm Fe(CN)}_6^{4-}]$, is
decreased~\cite{exp-spots}.  The homogeneous
system approaches a saddle--node bifurcation as
$[{\rm Fe(CN)}_6^{4-}]$ is decreased. Below a critical value there are three
fixed points (a low pH one, a high pH one, and an intermediate pH
saddle). Above this value only the low pH fixed point persists.
  We will show that both the patterns and this transition are
contained in the model family we propose.

The Gray--Scott and Fitzhugh--Nagumo models display
spot replication and lamellar patterns.
The Gray--Scott model is given by~\cite{john},~\cite{us}: 
\begin{eqnarray}
{{\partial u}\over{\partial t}}&=&D_u\nabla^2 u-
uv^2+A(1-u),\nonumber\\ {{\partial v}\over{\partial t}}&=&D_v\nabla^2
v+ uv^2-Bv,
\label{eq:rd_gs}
\end{eqnarray}
and it has been shown to behave similarly to the 
experiment~\cite{exp-spots}.
The Fitzhgugh--Nagumo model~\cite{aric},~\cite{muratov}, 
can be written as:
\begin{eqnarray}
{{\partial u}\over{\partial t}}&=&D_u\nabla^2 u
-\alpha(v+a_1u-a_0)
,\nonumber\\
{{\partial v}\over{\partial t}}&=&D_v\nabla^2 v+
v-v^3+u .
\label{eq:rd_fn}
\end{eqnarray}
The homogeneous dynamics of both models
is a flow in $R^2$. Furthermore, the families described by
Eqs.~(\ref{eq:rd_gs}) and (\ref{eq:rd_fn}) contain the flows of
Figs.~\ref{fig:exc}~(a)--(c). 
Thus, both systems have the ``right'' homogeneous dynamics to
\begin{figure}
\epsfxsize=1.6truein
\hskip -0.7 true cm \epsfbox{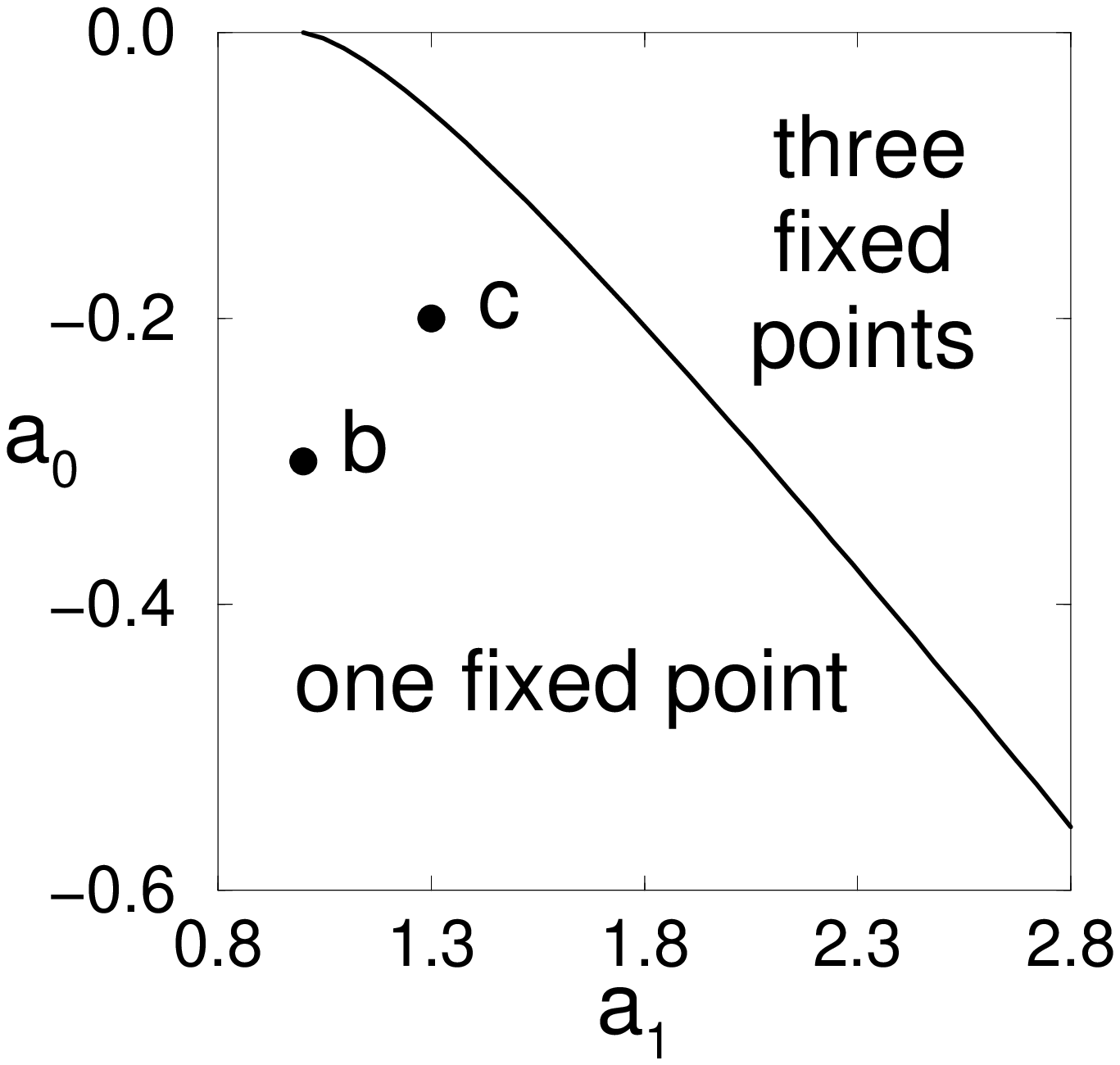}
\epsfxsize=1.3truein
\vskip -3 true cm\hskip  2.5 true cm\epsfbox{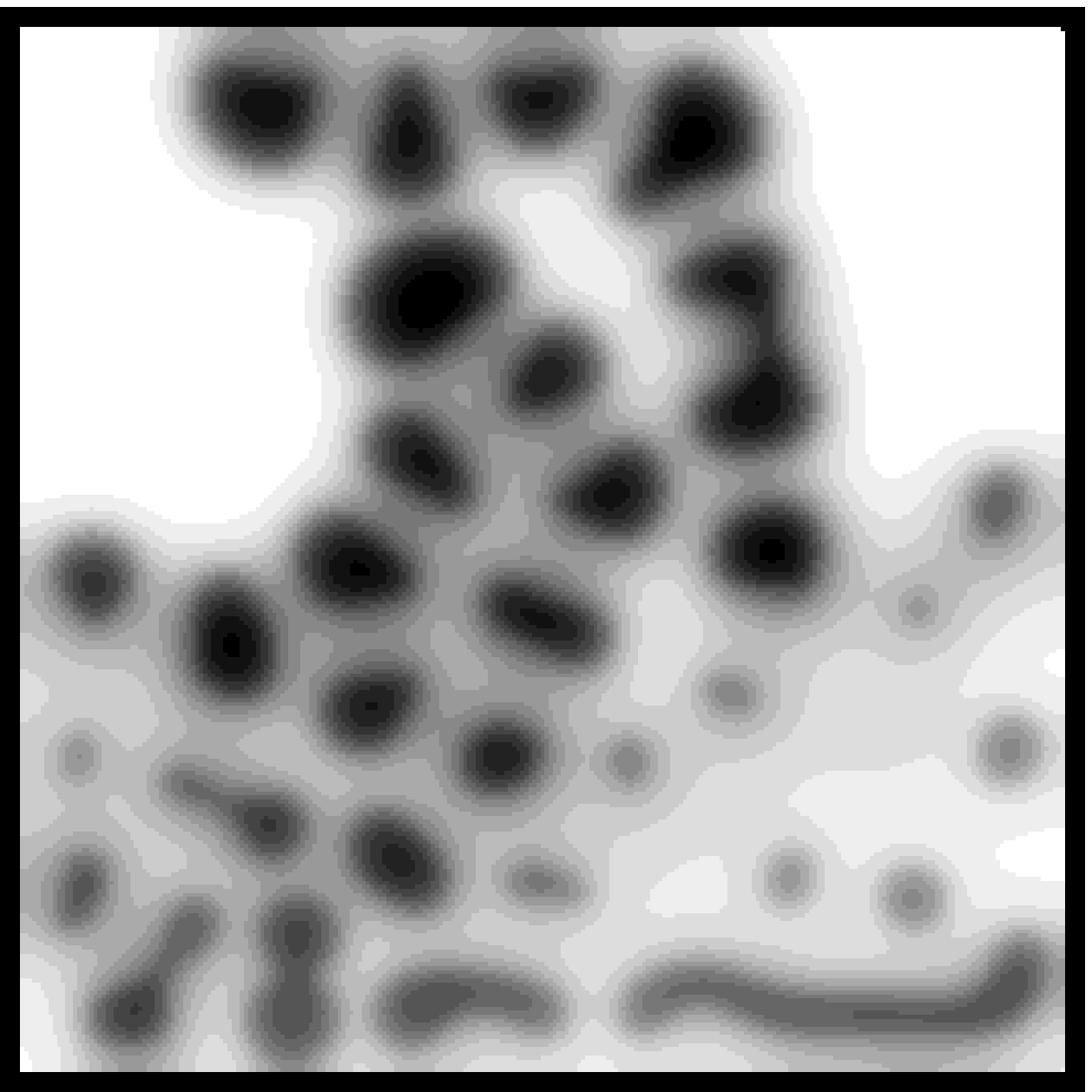}
\epsfxsize=1.3truein
\vskip -2.4 true cm\hskip  4.99 true cm\epsfbox{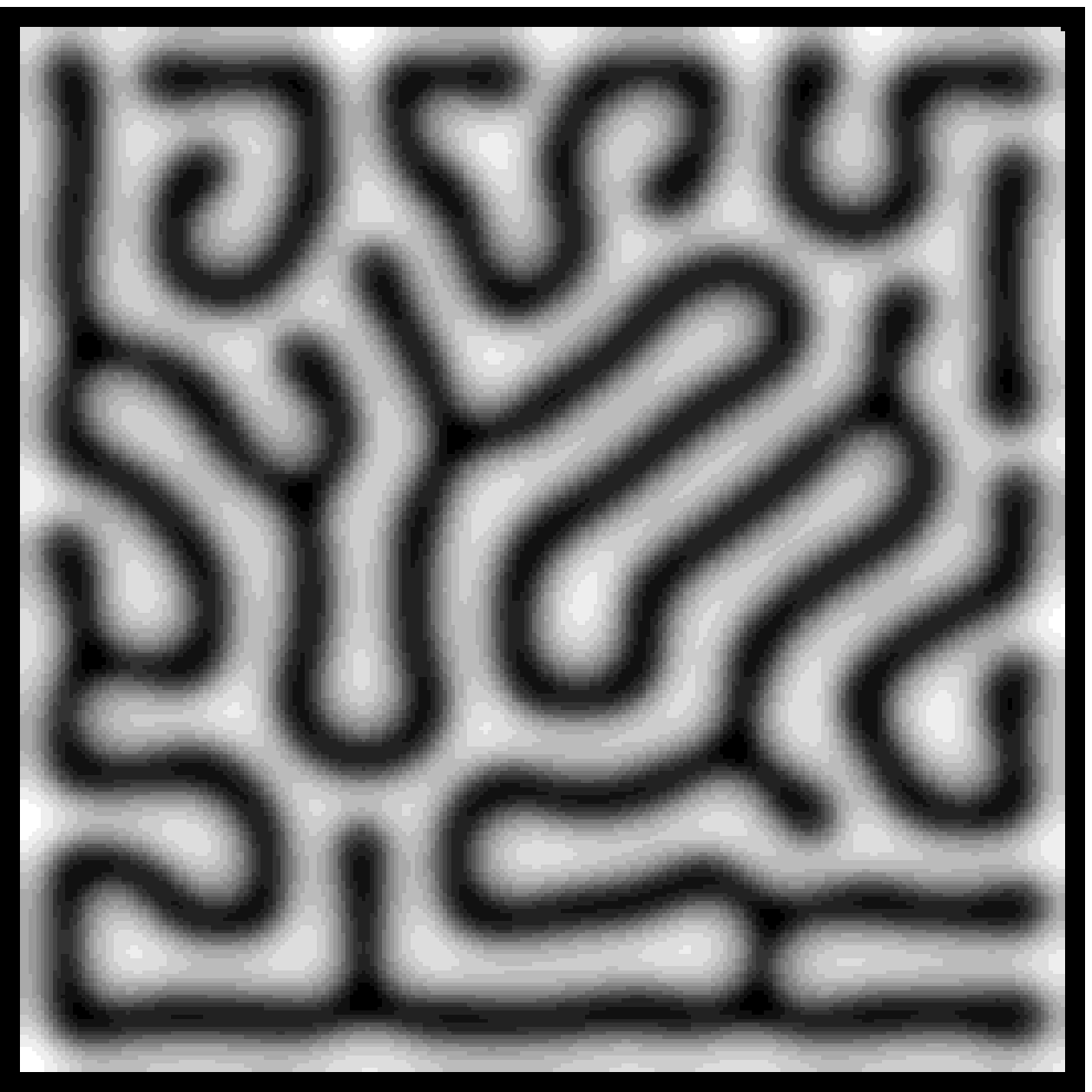}
%\end{figure}
\vskip -2.8 cm
\hskip 1.6 true cm
%\fontsize{14}{16pt}\selectfont
a
\hskip 2.2 cm
b
\hskip 2.2 cm
c
\vskip 3 true cm\caption{Similar to Fig.~\protect\ref{fig:gs}, but for the
Fitzhgugh--Nagumo model. The examples in (b) and (c) were obtained
for $D_u=1$ and $D_v=0.166$.
}
\label{fig:fn}
\end{figure}
construct model families of one another and of the FIS reaction. We
conclude that all these systems belong to the same equivalence
class. Furthermore, for the qualitative comparison we present, cross
diffusion terms are not necessary and either Eqs.~(\ref{eq:rd_gs}) or
(\ref{eq:rd_fn}) can be used as the model family for this class.

We show in
Figs.~\ref{fig:gs} and \ref{fig:fn} snapshots of some of the patterns
obtained in numerical simulations of Eqs.~(\ref{eq:rd_gs}) and
(\ref{eq:rd_fn}), respectively, when the systems have only one stable
but excitable fixed point.  We observe spot replication and
labyrinthine patterns.  As in the experimental system, we observe a
transition from spots to lamellar patterns as the homogeneous systems
approach the saddle--node (saddle--repellor) bifurcation (see the
partial bifurcation sets shown in Figs.~\ref{fig:gs}~(a) and
\ref{fig:fn}~(a))~\cite{diffusion}. We can provide a rough explanation for this if we
think of stationary solutions and regard diffusion as a perturbation
on the homogeneous dynamics at each spatial point.  In this sense,
diffusion makes the system cross the saddle--node (or
saddle--repellor) bifurcation. It is clear that the closer the
homogeneous system is to the bifurcation, the smaller must be the
perturbation and thus $\nabla^2 u$ and $\nabla^2 v$. For
this reason, more spatially extended patterns with smaller Laplacians
(such as lamellae) can be supported when the homogeneous system is closer to
the bifurcation point. 

We have proposed a global classification scheme for excitable reaction--diffusion
systems in terms of model families whose
choice is based on their underlying homogeneous dynamics. 
We have concluded
that the FIS reaction and Eqs.(\ref{eq:rd_gs}) and
(\ref{eq:rd_fn}), in the region of parameter space discussed, belong to the same equivalence
class.  This class is
organized around a Takens--Bogdanov point~\cite{jap} and we
call it the {\it Takens--Bogdanov model of excitability}. 
Families with other types of homogeneous dynamics will serve as templates for
other classes of excitable systems. 
We believe that this classification approach will  be possible whenever the
homogeneous dynamics can be mapped onto a planar flow.  The fact that
limit sets of planar flows are so restricted may be the explanation
of the ubiquitous presence of certain patterns in diverse
systems. For example, this might be the reason that the
complex Ginzburg--Landau equation reproduces behaviors outside its range of
applicability as a near--threshold normal form. 

This work was 
supported by the University of Buenos Aires, CONICET and Fundaci\'on
Antorchas and the Los Alamos National Laboratory LDRD program. 
We acknowledge useful conversations with G. Mindlin, C. Doering, and
B. Hasslacher. We would especially like to thank H.L. Swinney and
K.J. Lee for providing figure two.


\begin{thebibliography}{99}
\bibitem{aclar2} We also need to know how the flow behaves near
infinity, which can usually be regarded as a
collection of fixed points.
\bibitem{tak}
This normal form can either have a plus or a minus sign in front of the
$uv$ term (see~\cite{gh}).
\bibitem{aclar_pert}
In order to describe certain patterns it may suffice that
the model families contain only a proper subset of all 
perturbations. For the example discussed here the flow of 
Fig.~\ref{fig:exc}~(d) is unnecessary.
\bibitem{diffusion}
We note that
the patterns of Figs.~\ref{fig:gs}  and \ref{fig:fn}  were obtained for
$D_u/D_v>1$, 
with $u$ the inhibitor and
$v$ the activator in both cases. In the 2 variable model of the
experimental
system, ${\rm H}^+$ is the activator and HSO$_3^{-}$ is the inhibitor. 
As pointed out in~\cite{exp-spots}, a slowly diffusing
activator does not seem reasonable, since protons (${\rm H}^+$) diffuse
rapidly compared to the other species. However, the reduction from $10$
to $2$ variables in the FIS model kinetics involves replacing the
chemical steps $2 {\rm H}^+\rightarrow {\rm I}_2$, ${\rm I}_2+{\rm
HSO}_3^{-}\rightarrow 3
{\rm H}^+$ by the single one $2 {\rm H}^++{\rm HSO}_3^{-}\rightarrow 3
{\rm H}^+$. Thus, the activator of the 2 variable model has an effective
diffusion
coefficient that is determined
 by both the protons and iodine, I$_2$~\cite{future}. 
This effective diffusion coefficient may be small, since 
iodine  binds to the gel which, in turn,
substantially reduces its own effective diffusion coefficient.
\end{thebibliography}
\end{document}